# Study on A High-integrated Cloud-Based Customer Relationship Management System

Song He, Emre Erturk

## ABSTRACT

In the context of the business applications, integrating an on-premise customer Relationship Management (CRM) system with other systems used to be resource-consuming and complicated in terms of data and system interworking. With the help of cloud computing technology, the on-cloud business applications today have the capability of integrating CRM with ERP and other systems that allow large enterprises as well as small and medium companies (SMEs), in order to manage critical business processes, such as Sales, Marketing, Customer Service, Operations, Finance, Field Service and Project Service Automation, in a unified manner. Therefore, businesses can cover the entire customer lifecycle by efficiently exploiting all these resources. This case study analyses the on-demand CRM applications under the Microsoft Dynamics 365 product line. By studying the structures and how the Dynamics 365 CRM integrates with other applications, this paper provides a reference to companies that plan to switch their on-premise CRM the cloud-CRM.

**Keywords**: Cloud *Computing, Customer Relationship Management, Business Intelligence*

## 1. INTRODUCTION

As one source of business intelligence, using a customer relationship management (CRM) system is a critical strategy with multiple tools to enhance the profitability of enterprises by improving customers' and other key stakeholders' satisfaction and managing their relationships (Payne & Frow, 2005). However, the significant budget, long-term implications, and manpower demands of On-premise CRM solutions are unaffordable to small companies (IBM, 2013). Therefore, CRM vendors adopt new web-based and On-demand systems that allow their client to pay for the systems as a monthly fee (Özcanli, 2012). In recent years, the Customer Relationship Management (CRM) system has experienced significant changes. Certain aspects, such as the cloud-utilization, mobilization, socialization and artificial intelligence are the major tendencies of modern CRM (Chapalamadugu, 2018). Among these aspects, cloud computing technology is a critical factor due to the advantage of multi-systems integrating, which enable data to be shared across different cloud-based systems (Chapalamadugu, 2018). In the 1990s, the CRM had been considered as the link between the front office (sales, service, and marketing), back office (financial, operations, logistics, and human resources) and customers (Fickel, 1999). However, in the context of traditional CRM, companies are more likely to employ the CRM as a bridge between sales, marketing, and services (Chen & Popovich, 2003). This is because there were usually inconsistencies between various applications during the integration (Myron, 2003). Therefore, a seamlessly integrated, versatile and economic platform is highly desired by the users.

A traditional CRM (also known as On-premise CRM), is typically developed or hosted inside the company. To implement an On-premise CRM, the company needs to conduct a series of activities, such as deliverable planning, restructuring their sales and marketing teams, business process re-engineering, purchasing their won hardware and software, software customization, and training of employees (Özcanli, 2012). These activities can make up a significant expense. Hence the traditional CRM tends to be employed by large enterprises instead of SMEs, which would rather have a plug and play type of system (Özcanli, 2012). So, the companies have to seek new flexible and cost-effective solutions, one of which is a cloud-based CRM system.

In a cloud CRM, the system can perform tasks and store data online by means of utilizing the cloud computing infrastructure (TEC Team, 2018). This paper focuses on the CRM solutions delivered through the Software as a Service (SaaS) model. SaaS is now a mainstream approach to delivering software and outsourcing IT (Clark, 2016; Özcanli, 2012). This is also referred to as On-demand CRM as opposed to On-premise CRM. A typical On-demand CRM covers basic functions (Clark, 2016), such as sale force automation, marketing automation, customer service and support, analytics, reporting, and other extended CRM capabilities (TEC Team, 2018).

From the point view of CRM, there are several benefits, which cloud computing can provide. First is cost reduction. Companies do not need to spend a great deal of money in developing, purchasing, and maintaining a system. Second is scalability. The On-demand scalability allows companies to easily scale-up, scale-out, or scale-down their customer services (Wang et al., 2012). In addition, integration is another important benefit (Hwang, Dongarra, & Fox, 2013). According to Kumar and Petersen (2012), a key concern of Customer Relationship Management is integration with other business systems. Cloud computing can facilitate cross-organizational data and services (Iyer & Henderson, 2010). As for Big Data, on-cloud analysis also has a great advantage in terms of being able to scale up the data processing capacity (Schmidt & Möhring, 2013). Adopting these new information technologies is essential for New Zealand and other small but innovative economies, in order to maintain strong links with the rest of the world (Erturk & Fail, 2013).

This paper uses the cloud-based applications suite called Microsoft Dynamics 365, which includes a series of business applications and covers all necessary functions of CRM and ERP, including the common key business processes. Dynamics 365 for CRM covers sales, marketing, customer service and field service, while the ERP component in Dynamics 365 is also called "Finance and Operations", which includes numerous modules, such as finance, product management, inventory, and human resources. During this study, a Dynamics 365 account was set up to understand the platform. All the tests were done with the demo data set in Dynamics 365. Firstly, this paper reviews Dynamics 365 generally, in terms of its structure. Later, the paper studies the integration capabilities of Dynamics 365, according to three dimensions, including business applications, productivity tools, and intelligence tools. In the last section, the technologies and tools that are used for integration are reviewed and discussed.

## 2. CASE STUDY
### 2.1 Dynamics 365 Overview
Dynamics 365 is a cloud-based solution that combines CRM and ERP applications together into one platform and allows users to personalize their own business processes or applications consistently within a common data model. Also, this platform is tied together with various productivity tools, such as Office365 and Power BI, which are familiar to users in the context of business processes. Meanwhile, the inbuilt intelligence module not only provides the visualized analytics on a dashboard but can also work efficiently through the interactive automation functions. Furthermore, this platform is extensible, which means users can leverage many other software-as-a-service (SaaS) extensions available as Microsoft apps.

Figure 1 is the Dynamics 365 in Microsoft Cloud framework, which is a highly integrated ecosystem. Microsoft Azure is the platform that hosts the entire suite of cloud applications and the common data model. Above the platform, cloud-based applications and services, such as Office 365, Power BI, Cortana Intelligence, Azure IoT, Dynamics 365, and the other third-party business applications, make up the whole business application layer. The upper layer of the framework is a central hub of applications called Microsoft AppSource, in which everyone can develop and publish their own business apps.

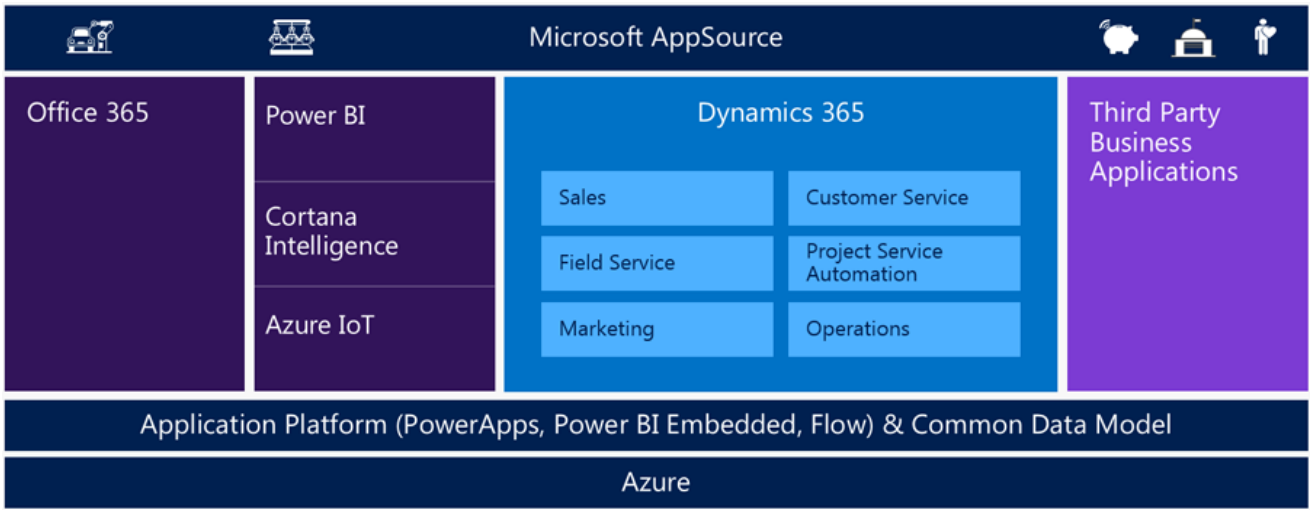

*Figure 1: Dynamics 365 in Microsoft Cloud framework. (from Chorus, 2016)*

### 2.2 Business Applications Integration
From the overview of Dynamics 365 for CRM in Figure 2, all the available CRM modules are gathered under the drop-down menu which allows users to navigate to the necessary process in a system-wide manner. Likewise, different CRM modules have cross functions with other modules. For example, in Figure 2, the Marketing Lists and Quick Campaigns under Marketing module exist in Sales, which enable user from the sales department to have a quick view as to what the customer likes or what channels the lead was collected from during the marketing campaign activities.

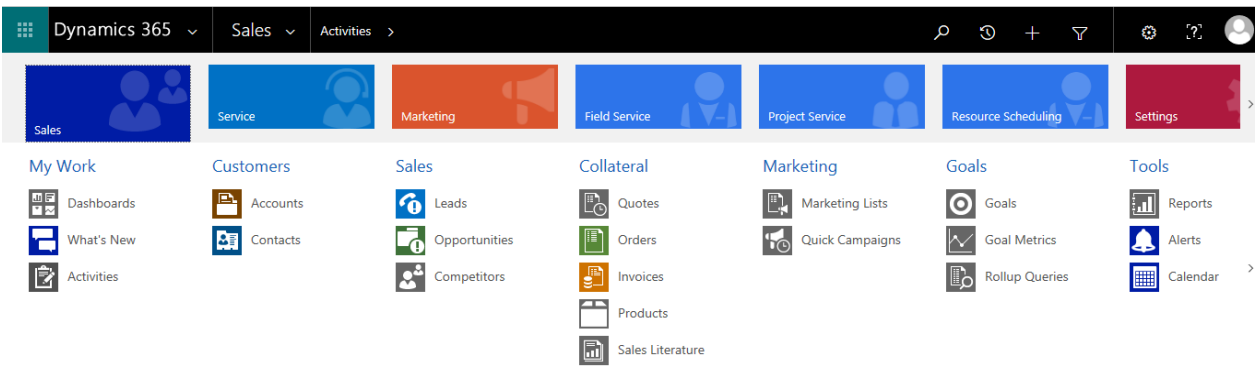

*Figure 2: Dynamics 365 for CRM user interface overview (2018)*

The business application integration focuses on the interaction between CRM and ERP. This study selects some data components to get a better understanding of how applications under CRM and ERP work with each other. Figure 3 shows the marketing leads data in Dynamics 365, in which marketing and sales applications share the same Leads module. This is because workflows and campaign activities processed in the marketing department generate or involve leads, while follow-up processes including qualify, develop, propose and close, are conducted in the sales department. Sharing leads data is necessary for the sales application and the marketing application to be well integrated.

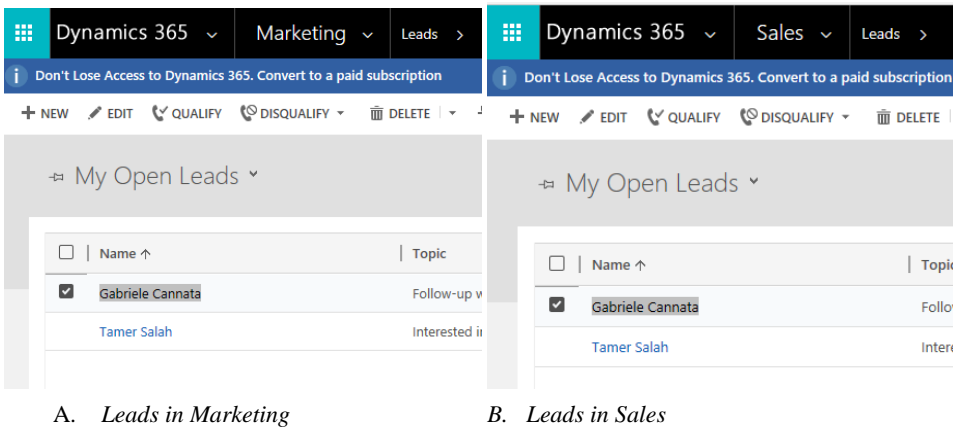

| A. Leads in Marketing | B. Leads in Sales |

*Figure 3: Leads Data in Dynamics 365 (2018)*

After qualifying the first lead (the "Gabriele Cannata"), the system automatically generates a new opportunity involving an account and a contact based on the information of this lead. Figure 3 is this contacts data in different applications under Dynamics 365. In Figures A and D, the contacts data is the same. This indicates that the same contacts data are shared by sales, service, marketing, and field service application.

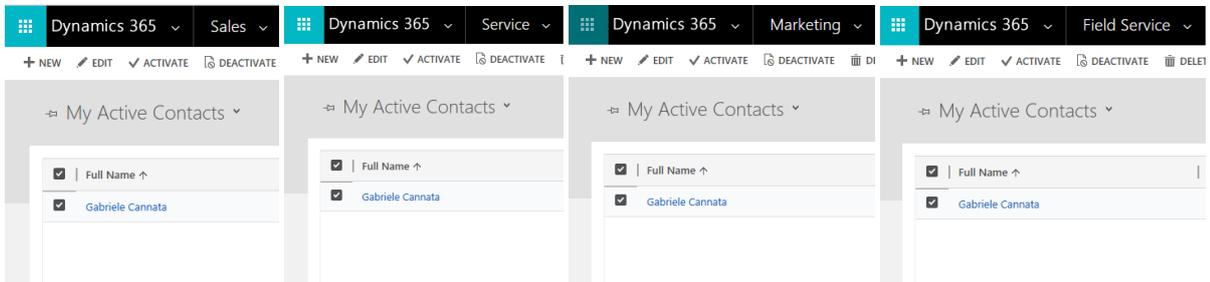

A. Contacts in Marketing    B. Contacts in Sales    C. Contacts in Marketing    D. Contacts in Field Service

*Figure 2: Contacts Data in Dynamics 365 (2018)*

Regarding the interconnection between CRM and ERP, since the study does not subscribe to a pricing plan which would have the full functionality, the CRM and ERP applications are separately utilized and do not have any data interchange. However, this study compares the modules of applications against CRM and ERP, and finds several modules commonly existing in these applications. Figure 4 shows the sales and marketing modules group of Dynamics Finance and Operations, in which the data of leads and contacts are also included. These basic data modules are applied to the back-office finance and operation processes involved in invoices and products. In addition, key CRM business processes, such as opportunities, quotations, and sales orders, also exist in ERP. This is because these processes include data points of Finance and Products.

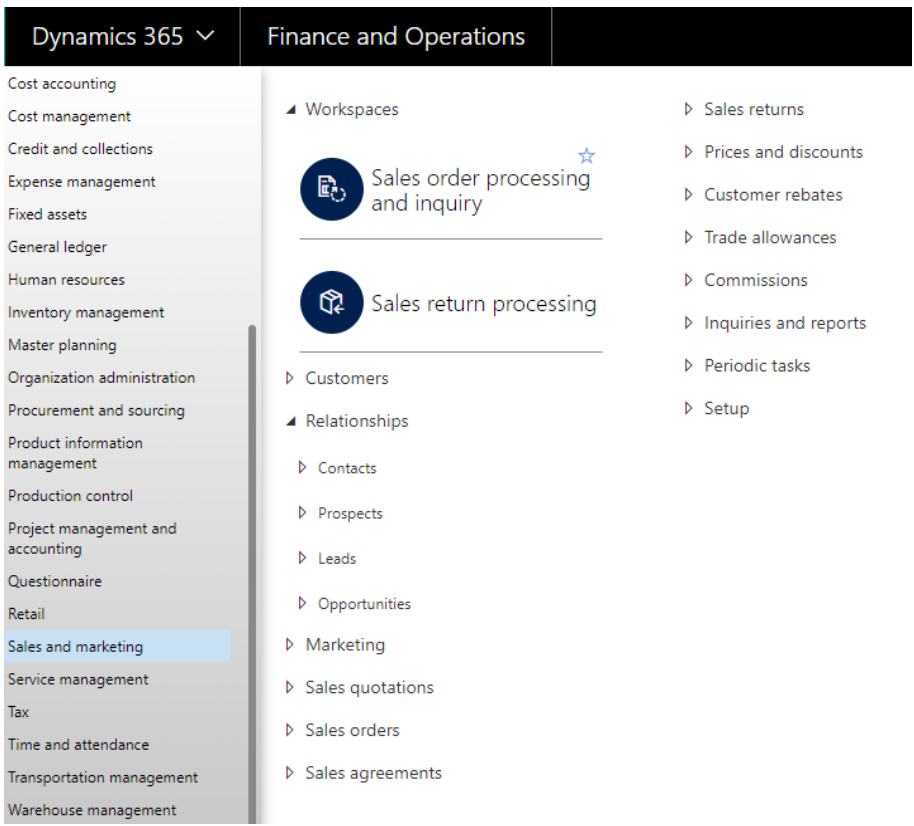

*Figure 4: Sales and Marketing Modules of Dynamics Finance and Operations (2018)*

## 2.3 Productivity-Business Integration

By leveraging the cloud platform, Dynamics 365 for CRM can easily collaborate with productivity tools, such as Office Online and Skype. To showcase the integration of these tools, this paper tested two functions: exporting data and online dialing.

Figure 5 is the opportunities data list under the Sales module. When the user selects the "Open in Excel online", the system will execute an online Excel module as in Figure 6. Users can edit and save this document in an online directory synchronized with the Microsoft OneDrive for Business cloud-based drive. By leveraging this integration, users can keep files completely in the cloud and can run tasks on multiple devices.

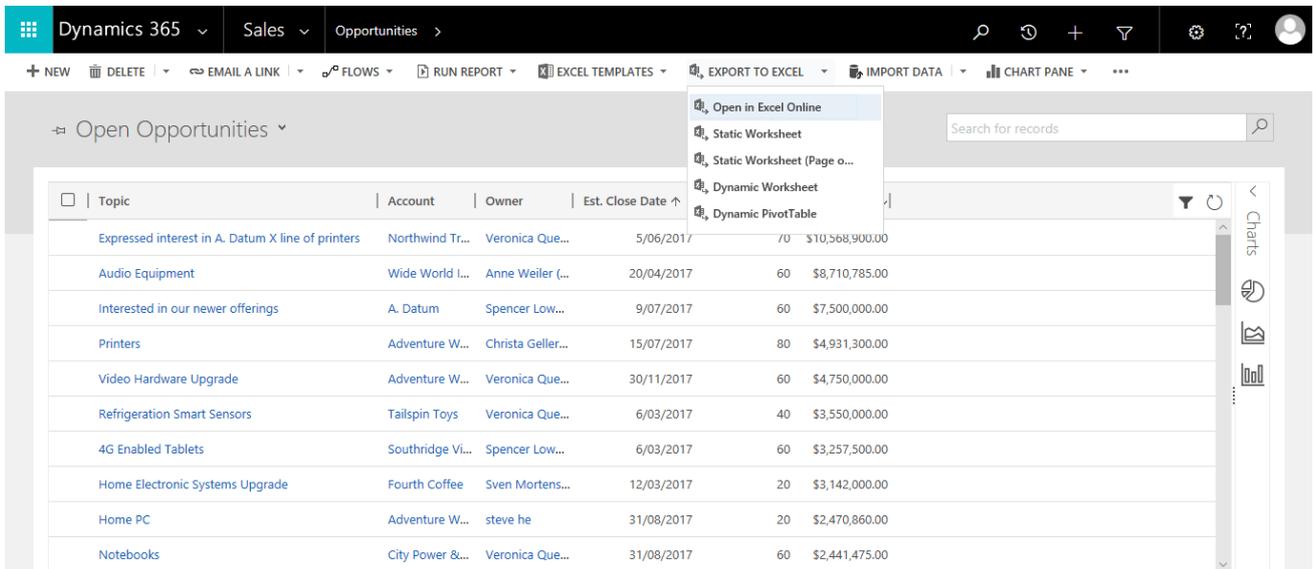

*Figure 5: Opportunities List of Dynamics 365 (2018)*

*Figure 6: Opportunities List in Online Excel (2018)*

Connecting with customers is a critical work task. Dynamics 365 has seamlessly integrated Skype for communications. In this example, when the user clicks the number on the screen (see Figure 6), a Skype window pops up instantly (see Figure 7). This integration enables a sales person to conduct a one-click-dial telemarketing, provided Skype is installed on their computer.

*Figure 7: Customer Detail (2018)*

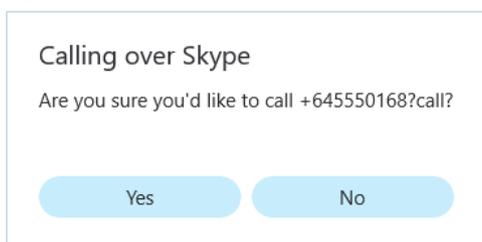

*Figure 8: Skype Application Popup Window (2018)*

## 2.4 Intelligence Integration

To understand the intelligence integration, this study tried to test the Azure Machine Learning in Dynamics 365. During the sales process, the system has a product recommendation function, which employs this technology. Figure 9 is the product suggestion window from a customer demo data. Part A of Figure 9 shows the detail of the product the customer is buying, while Parts B, C, and D show the suggestions. In this scenario, the products are Carbon Fiber 3D Printers, of which the suggestions function window shows on up-sell item, three cross-sell items, and zero accessories and substitutes in Parts B, C, and D.

According to Microsoft TechNet (2017), in the context of automation and machine learning, the recommendation modelling techniques can automatically suggest products based on the behavioural data. In addition, "product recommendations feature supports existing line item entities (Opportunity Product, Quote Detail, Sales Order Detail, and Invoice Detail) and custom line item entities, as well as standard and custom product relationships" (Microsoft TechNet, 2017, para.5).

*Part A. Customer's Product Line Items*

*Part B. Cross Sell*

*Part C. Accessory*

*Part D. Up Sell*

*Part E. Substitute*

*Figure 9 "Products Suggestions Function" (2018)*

## 3. TECHNOLOGIES AND DISCUSSION

Dynamics 365 is an example of a SaaS cloud platform, in which all Microsoft applications can be easily accessed from anywhere (Turnkey Technologies, 2016). Likewise, these applications are SaaS-based, which allow users to subscribe at their own business scale.

Regarding the integration, according to Iyer and Henderson (2010), cloud computing technology has an advantage in integrating data and service across different organizations. In Dynamics 365, with the underlying Azure Cloud Service, all the SaaS applications can be gathered into one platform. In addition, to maintain the information flow between CRM and ERP, Dynamics 365 uses a consistent data model named Common Data Service (CDS), in which the most common data entities are shared by all the connected applications (Cole, 2017). Furthermore, by leveraging AppSource, Dynamics 365 can connect numerous third-party applications. This factor can make Dynamics 365 more scalable and powerful than other CRM systems that rely on the third-party plug-ins (Chorus, 2016).

Although Microsoft Dynamics 365 has taken good advantage of systems integration, there are also several challenges, related to the sophistication of the modules. The first issue is performance. During the study, the forms' opening time was unstable, despite the simple demo data load. The performance issue may have been partly caused by the network speed of the academic test environment. However, this aspect should be further studied and addressed because adoption in academic and public environments may often be quite important for the long-term success of a software technology (Erturk, 2009). Anyhow, in the context of an industrial or commercial implementation, performance will rely largely on the available network and server resources. Other factors that influence performance include the capabilities of client devices and the particular Dynamics 365 configuration being used (Webb, 2017).

The second challenge has to do with fast-changing features, i.e. short release cycles of Dynamics 365. From January 2016 to May 2018, Dynamics 365 for CRM (named Microsoft Dynamics CRM Online before version 8.1) had experienced 93 releases and updates (Microsoft Support, n.d.). Although regular updates can improve the system's capability and security, bring better experiences to users, the highly frequent releases are likely to also cause confusion to users, and confusing them while learning and keeping up with the changes (Lindstrom, 2017). An alternative scenario is for Microsoft to perform a silent update, which may go unnoticed by the users (Plourde & Horwath, 2016).

## 4. CONCLUSION

In general, Dynamics 365 CRM is a typical cloud-based CRM involving various SaaS applications and facilitating multiple business practices. From the perspective of the system or platform, the entire Dynamics 365 is a cloud-based ecosystem, in which the CRM, ERP, Intelligence, and third-party applications are organically integrated.

In this paper, integration was examined from three dimensions. Firstly, the integration of business applications was studied by means of checking the data sharing, including the leads and sharing among applications within CRM and between CRM and ERP. As a result, this paper found that specific business processes, such as opportunities, quotations, and sales orders also existed in both CRM and ERP.

Next testing was the productive-tools integration of Dynamics 365 CRM. By studying the export functions, this paper found that Dynamics 365 CRM worked well with Excel Online and can provide a cloud document storage service. Besides, the study also examined Skype integration by testing the one-click dialing function, which enables CRM users to connect with contacts via Skype by clicking the phone number's link under their profile.

The third aspect is intelligence integration. According to the Microsoft support documentation, this functionality is provided in Dynamics 365 through the Azure platform.

Although Dynamics 365 offers many integration advantages through the single cloud-based platform, the system performance stability and frequent updates add complexities to this CRM system. We can anticipate that Microsoft will continue to address these challenges.